\documentclass[sigconf,screen,nonacm]{acmart}  %

\AtBeginDocument{%
  }

\copyrightyear{2025}
\acmYear{2025}
\setcopyright{none}
\acmConference[E-ENERGY '25]{The 16th ACM International Conference on Future and Sustainable Energy Systems}{June 17--20, 2025}{Rotterdam, Netherlands}
\acmBooktitle{The 16th ACM International Conference on Future and Sustainable Energy Systems (E-ENERGY '25), June 17--20, 2025, Rotterdam, Netherlands}
\acmDOI{10.1145/3679240.3734614}
\acmISBN{979-8-4007-1125-1/2025/06}

\usepackage{booktabs}
\usepackage{tablefootnote}
\usepackage{multirow}

\usepackage{algorithm}
\usepackage[noend]{algpseudocode}

\usepackage{enumerate}

\newcommand{\LineComment}[1]{\State {\small\color{teal} \# #1}}

\newcommand{\parag}[1]{\vspace{1mm plus 1mm}\noindent\textbf{#1.}\hspace{1mm}}

\newtheorem{assumption}{Assumption}
\newtheorem{theorem}{Theorem}
\newtheorem{lemma}{Lemma}

\usepackage{todonotes}

\newcommand{\QoRtarget}{\text{QoR}_\text{target}}

\newcommand\blfootnote[1]{%
  \begingroup
  \renewcommand\thefootnote{}\footnote{#1}%
  \addtocounter{footnote}{-1}%
  \endgroup
}

\usepackage{subcaption} %

\begin{document}

\title{Carbon-Aware Quality Adaptation for~Energy-Intensive Services}

\author{Philipp Wiesner}
\orcid{0000-0001-5352-7525}
\affiliation{%
  \institution{Technische Universität Berlin}
  \city{Berlin}
  \country{Germany}
}
\email{wiesner@tu-berlin.de}

\author{Dennis Grinwald}
\orcid{0000-0001-9903-2886}
\affiliation{%
  \institution{Technische Universität Berlin \& BIFOLD}
  \city{Berlin}
  \country{Germany}
}
\email{dennis.grinwald@tu-berlin.de}

\author{Philipp Weiß}
\orcid{0009-0003-7228-5076}
\affiliation{%
  \institution{Technische Universität Berlin \& BIFOLD}
  \city{Berlin}
  \country{Germany}
}
\email{weiss@tu-berlin.de}

\author{Patrick Wilhelm}
\orcid{0009-0004-8714-5647}
\affiliation{%
  \institution{Technische Universität Berlin \& BIFOLD}
  \city{Berlin}
  \country{Germany}
}
\email{patrick.wilhelm@tu-berlin.de}

\author{Ramin Khalili}
\orcid{0000-0003-2463-7033}
\affiliation{%
  \institution{Huawei Technologies}
  \city{Munich}
  \country{Germany}
}
\email{ramin.khalili@huawei.com}

\author{Odej Kao}
\orcid{0000-0001-6454-6799}
\affiliation{%
  \institution{Technische Universität Berlin}
  \city{Berlin}
  \country{Germany}
}
\email{odej.kao@tu-berlin.de}

\renewcommand{\shortauthors}{Wiesner et al.}

\begin{abstract}
The energy demand of modern cloud services, particularly those related to generative AI, is increasing at an unprecedented pace.
To date, carbon-aware computing strategies have primarily focused on batch process scheduling or geo-distributed load balancing.
However, such approaches are not applicable to services that require constant availability at specific locations due to latency, privacy, data, or infrastructure constraints.

In this paper, we explore how the carbon footprint of energy-intensive services can be reduced by adjusting the fraction of requests served by different service quality tiers.
We show that adapting this quality of responses with respect to grid carbon intensity can lead to additional carbon savings beyond resource and energy efficiency.
Building on this, we introduce a forecast-based multi-horizon optimization that reaches close-to-optimal carbon savings and is able to automatically adapt service quality for best-effort users to stay within an annual carbon budget.
Our approach can reduce the emissions of large-scale LLM services, which we estimate at multiple 10,000 tons of CO$_2$ annually, by up to 10\,\%.
\end{abstract}



\maketitle

\blfootnote{This manuscript is an extended version of our paper published at \textbf{ACM e-Energy'25} (DOI: \href{https://dl.acm.org/doi/10.1145/3679240.3734614}{10.1145/3679240.3734614}). Compared to the published version, we
\begin{enumerate}
    \item add a time-based vs.\ utilization-based power attribution perspective together with a proof that both yield equivalent provisioning decisions under mild assumptions (Appendix~\ref{sec:proof_eq}),
    \item extend the online approach with an automatic adaptation of $\QoRtarget$ to meet a fixed annual carbon budget (Sections~\ref{sec:automatic_qor_budget} and \ref{sec:eval_budget}).
\end{enumerate}
}

\section{Introduction}

Operating modern cloud services can require substantial computing resources and energy, which directly contributes to the rapidly growing carbon footprint of computing systems~\cite{snp2024datacenter_trends}.
Alongside traditional resource-intensive services like real-time data analytics, video processing, recommendation systems, and streaming, the growing use and high energy demand of generative AI is currently driving the expansion of data centers at a high pace~\cite{Bashir2024Climate, wu2024scalingAIsustainably}. %
As a result, major cloud providers, who collectively pledged to reduce their carbon footprint to zero by 2030, have been increasing their emissions significantly in recent years~\cite{Google2024EnvReport, Microsoft2024EnvReport, meta2023sustainability} and are expected to row back on sustainability commitments~\cite{Lawrence2024NetZero}.

The concept of carbon-aware computing has emerged as a response to such challenges:
Carbon-aware strategies aim to reduce emissions beyond energy efficiency by additionally taking into account how carbon-intensive the consumption of electricity at a current time and location is.
Especially in the context of scheduling flexible batch workloads, the potential of temporal and spatial load shifting is well understood~\cite{wiesner2021letswaitawhile, Radovanovic_Google_2022, Zheng_MitigatingCurtailment_2020, wiesner2024fedzero,Hanafy_GoingGreenLessGreen_2024, sukprasert2024eurosys, carbonscaler2024sigmetrics}.
In contrast, existing carbon-aware solutions for interactive services, which handle individual requests with minimal delay-tolerance, predominantly focus on geographical load balancing~\cite{gsteiger2024caribou, murillo2024cdn_shifter, souza2023casper, Zhou2016Carbon-Aware, Zhou2013Carbon-Aware}.
However, many services cannot simply be distributed across regions due to data locality requirements, regulatory constraints, privacy and security concerns, or infrastructure limitations.
Carbon-aware approaches for services that are constrained to a single region remain largely unexplored.

\begin{figure}
\centering
  \includegraphics[width=.99\columnwidth]{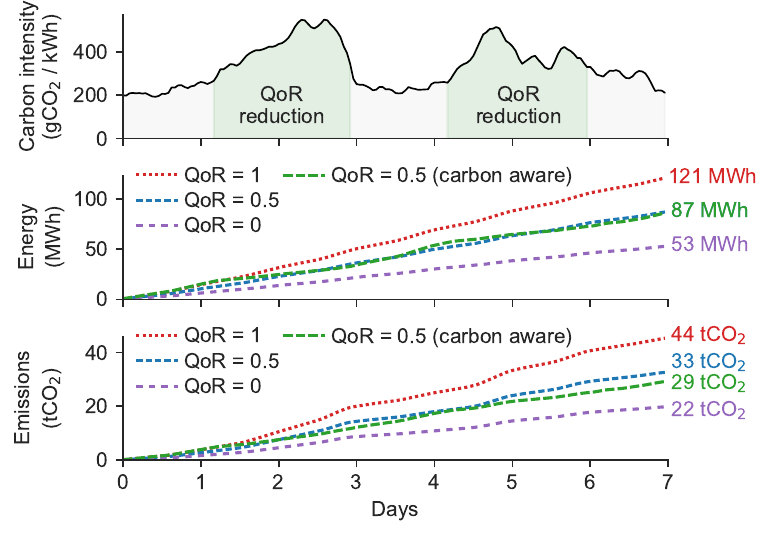}
  \vspace{-4mm}
  \caption{We investigate how adjusting the quality of responses (QoR) with respect to carbon intensity can reduce emissions beyond savings related to reduced energy demand.}
  \label{fig:figure1}
\end{figure}

In this work, we focus on the \emph{temporal} variability of carbon intensity and explain how adapting service quality with respect to this metric does not only impact energy demand, but can yield additional carbon savings.
We define the quality of responses~(QoR) as a type of quality of service~(QoS) metric that describes the proportion of requests served by two different service quality tiers.
For instance, the service quality and carbon footprint of LLM inference is determined by many factors such as the choice of model~\cite{faiz2024llmcarbon}, quantization and pruning~\cite{liang2021pruning_quantization}, and the number of generated tokens~\cite{li2024sprout}.

Figure~\ref{fig:figure1} shows the accumulated energy usage and carbon emissions for an example scenario with two quality tiers over one week.
Here, serving all requests at the high quality tier (QoR = 1) uses roughly twice the energy compared to serving them at the low quality tier (QoR = 0).
Balancing requests evenly between the two tiers results in intermediate energy usage.
We explore how strategically timing quality reductions based on carbon intensity can further reduce emissions by performing a comprehensive analysis across different regions, request patterns, and QoR constraints.

\parag{Contributions}
Towards more carbon-efficient cloud services, we make the following contributions:

\begin{enumerate}
    \item We formalize the problem of minimizing emissions under QoR constraints through optimized resource provisioning and load balancing between two service quality tiers.
    \item We propose a forecast-based multi-horizon optimization for minimizing emissions that reaches close-to-optimal carbon savings.
    \item We evaluate the potential and practicability of carbon-aware QoR adaptation by simulating a large-scale LLM service.
\end{enumerate}
All code and data is publicly available\footnote{\url{https://github.com/dos-group/qualitytime}}.

\section{Carbon-Aware QoR Adaptation}

Offering different service quality tiers is common practice in the operation of cloud services.
For example, content delivery networks provide faster and more reliable service to higher-tier users, while AI platforms like OpenAI and Gemini reserve access to larger and more advanced models for paying customers.
Furthermore, service providers already degrade service quality during periods of high load to prevent system overload~\cite{park2024topfull, meta2023defcon}.
Such measures often target best-effort users: For instance, free-tier users of LLM inference services can experience highly increased latencies at certain times, to not endanger the QoS of subscribed users.

Recently, there have been proposals to design applications that can intentionally sacrifice QoS, not only to manage economical interests or load peaks, but as a strategy to improve sustainability~\cite{vitali2022sadp, vitali2023enriching_cloud_with_sustaimnability_ic2e}.
In this paper, we follow the predominant assumption that future regulations for Scope~2 emission reporting will be based on average carbon intensity (ACI), reported in carbon-equivalent emissions per unit of energy (gCO$_2$/kWh)~\cite{gsteiger2024caribou, wiesner2021letswaitawhile, Hanafy_GoingGreenLessGreen_2024, sukprasert2024eurosys, Radovanovic_Google_2022}.
We examine the impact of dynamically adjusting service QoR with respect to temporal changes in ACI to enhance carbon efficiency without relying on geo-distributed load balancing.

\parag{Problem setting}
Given an interval duration~$\Delta$, we divide our optimization time window $[0,T]$ into $I = T/\Delta$ time intervals, indexed by $i$.
For the remainder of this paper, we consider $T = 1~\text{year}$ and $\Delta = 1$ hour.
We define the set of user groups as $\mathcal{U}$, machine types $\mathcal{M}$, and a tuple of two service quality tiers $\mathcal{Q} = (\text{Tier~1},\text{Tier~2})$.
Machine types can represent physical hardware as well as virtual machines (VMs) like cloud instances.
For each time interval, there is an associated carbon intensity $C^i$ and a number of requests $R^i$ per user group.
Each machine can handle a certain number of requests by serving a certain quality tier (denoted as $K$), which corresponds to its current power usage $P^i$.
The deployment $D^i$ must be sufficient to meet the number of incoming requests that are allocated to different service quality tiers ($A^i$). %
The key challenge lies is determining the optimal values for both $D^i$ and $A^i$ for each interval.
Appendix~\ref{sec:nomenclature} shows an overview of all input and decision variables.

\parag{Quality of Responses}
We define $\text{QoR} \in [0,1]$ as a metric that describes the proportion of requests served by two service quality tiers.
That is, $\text{QoR} = 0$ indicates that all requests are served by the lower-quality Tier~1, while $\text{QoR} = 1$ indicates that all requests are served by the higher-quality but more energy-intensive Tier~2.
When $\text{QoR} = 0.5$, half of the requests are served by Tier~1 and half by Tier~2.
Formally, for any interval $\bigcup_{i=\alpha,\dotsc,\omega} i$, QoR is defined as
\begin{align}
\text{QoR}(\alpha, \omega) = \frac{\sum_{i=\alpha}^{\omega} a^i_{u,q}}{\sum_{i=\alpha}^{\omega} r^i_u},
\end{align}

\begin{figure}
\centering
  \includegraphics[width=.94\columnwidth]{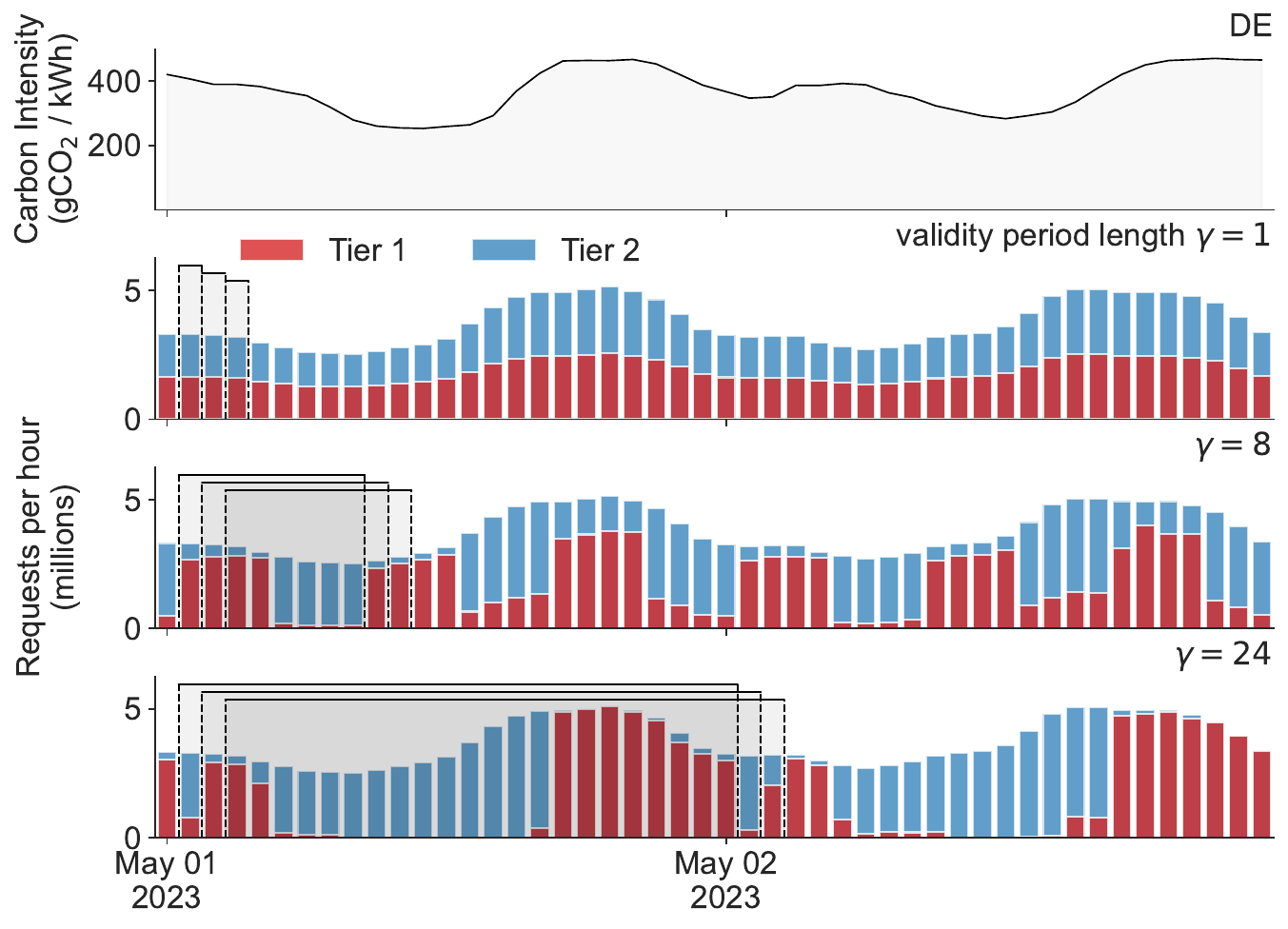}
  \vspace{-2mm}
  \caption{Increasing validity periods allow for more flexibility in adjusting the proportion of requests served by Tier 1 and Tier 2 based on carbon intensity.}
  \label{fig:validity_periods}
  \vspace{-2mm}
\end{figure}

\parag{Validity periods}
As is typical in the assessment of QoS metrics~\cite{keller2003WSLAFramework}, QoR is defined over validity periods of fixed length.
These periods can either be disjoint (e.g., quarters, months, weeks) or, more commonly, overlapping.
For the remainder of this paper, we assume that QoR is assessed over rolling windows of length $\gamma$ (integer multiple of adjustment interval $\Delta$).
Staying with the above example, when the validity period is set to $\gamma = 24~\text{h}$ and we want to guarantee a specific $\QoRtarget = 0.5$, we have to ensure that within each 24 hour window at least 50\,\% of all requests are served by Tier 2.
Formally, $0.5 \leq \min \left\{ \text{QoR}(i, i + \gamma)\right\}_{i = \alpha}^{\omega}$.
Figure \ref{fig:validity_periods} illustrates how the validity period length influences the hourly QoR over time.
Longer validity periods provide greater flexibility in balancing service quality in response to fluctuations in carbon intensity, but can lead to prolonged periods of high or low hourly QoR.

\parag{Carbon emission model}
We define a service's estimated carbon emissions $E^i$ during time interval $i$ with respect to the deployments current power usage, the region's carbon intensity $C^i$ (in gCO$_2$/kWh), and the attributed embodied emissions $C_m^\text{ emb}$ (in gCO$_2$) for running machine $m$ for time $\Delta$:
\begin{align}
\label{eq:emissions} E^i = \sum_{m \in \mathcal{M}} \sum_{q \in \mathcal{Q}} d^{i}_{m,q} \ \left( \Delta p^i_{m,q} \ C^i + C_m^\text{emb} \right)
\end{align}
Depending on the methodology, embodied emissions may be excluded from scheduling decisions~\cite{bashir2024sunkcarbon}, in which case $C_m^\text{emb} = 0$.
We do not explicitly model emissions per user or request in this paper, but they can be derived using carbon attribution models~\cite{HotCarbon2024GameTheoretic, Westerhof2023AllocationModel}.

\parag{Time-based vs.\ utilization-based power attribution}
The operational emissions in \eqref{eq:emissions} require a choice of how to attribute power draw to a service and its resources.
Two common modeling choices are:
\emph{(i) utilization-based} power attribution, where per-machine power depends on resource utilization, and \emph{(ii) time-based} power attribution, where every active machine is assigned a (typically constant) attributed power draw per unit time.

Utilization-based models are a natural fit when modeling operator-side electricity usage (often reported as Scope~2), where metering reflects the actual energy consumed by servers.
Time-based attribution is common in tenant-side carbon accounting (often reported as Scope~3) when emissions are allocated proportional to instance-hours.
However, Scope~3 methodologies can also be utilization- or energy-based, and Scope~2 reporting can rely on attributed values depending on available measurements.

For utilization-based attribution, machine power draw can be modeled as utilization-dependent, e.g., with a concave power law,
\begin{align}
\label{eq:scope2_powermodel_simple}
p^i_{q} = p^\text{idle} + \bigl(p_{q}^\text{max} - p^\text{idle}\bigr) \cdot \bigl(\text{util}^i_q\bigr)^n, \quad n \in (0,1],
\end{align}
where, in our setting with one user group and one machine type, $\text{util}^i_q = a^i_{q} / (d^i_{q} \cdot k_{q})$.
Under time-based attribution, $p^i_{q}$ is treated as utilization-independent (constant per active instance).
We show that under mild concavity assumptions, optimizing under time-based vs.\ utilization-based attribution yields equivalent provisioning decisions (Appendix~\ref{sec:proof_eq}).

\parag{Optimization problem}
We formalize the problem of optimizing service deployments $D^i$ that minimize carbon emissions, while providing an allocation $A^i$ of requests to different service quality tiers that satisfies a $\QoRtarget \in [0,1]$:
\begin{align}
    \min \quad
    \label{eq:objective_1} & \sum_{i=\alpha}^\omega E^i \\
    \text{s.t.}\quad
    \label{eq:all_requests_allocated} &\sum_{q \in \mathcal{Q}} a^{i}_{u,q} = r^{i}_{u} & \forall i, \forall u \\
    \label{eq:sufficient_resources} &\sum_{u \in \mathcal{U}} a^{i}_{u,q} \leq \sum_{m \in \mathcal{M}} d^{i}_{m,q} \cdot k_{m,q} & \forall i, \forall q \\
    \label{eq:min_qos_constr} & \QoRtarget \leq \min \left\{ \text{QoR}(i, i + \gamma)\right\}_{i = \alpha}^{\omega}
\end{align}
where \eqref{eq:all_requests_allocated} ensures that all requests during each interval $i$ are attributed to a quality tier, \eqref{eq:sufficient_resources} ensures that the provided machines have sufficient capacity to serve all requests per quality tier, and \eqref{eq:min_qos_constr} ensures that the QoR of every validity period stays above $\QoRtarget$.

Note, that this problem is not only NP-hard (see Appendix~\ref{sec:nphard}) but also assumes knowledge of future carbon intensity~($C^i$) and requests~($R^i$).
As such, it cannot be solved directly in practice.

\section{Online Approach}
\label{sec:online}

A practical approach for carbon-aware QoR adaptation must cope with two sources of inaccuracy:
\begin{enumerate}
    \item \textbf{Approximation errors}: Optimally solving the MILP over a large solution space is computationally infeasible. In our experiments, no annual-horizon instance solved within 1 hour, while the median runtime on daily-horizon instances was 1.2 seconds (see Appendix~\ref{sec:performance_analysis}).
    \item \textbf{Forecast errors}: Long-term forecasts tend to be less accurate than short-term ones, particularly in regions with volatile renewable generation~\cite{maji2022carboncast, li2024uncertaintyAwareCarbonDatacenters}.
\end{enumerate}

Based on these findings, we propose a multi-horizon optimization which periodically updates forecasts and determines an optimal (or close-to-optimal) deployment and load balancing for every interval $i$, described in Algorithm~\ref{alg:1}.

\begin{algorithm}[h]
    \caption{Multi-Horizon Optimization}
    \algrenewcommand\algorithmicindent{1.0em}
    \label{alg:1}
    \begin{algorithmic}[1] %
    \For{$\alpha \gets 0$ \textbf{to} $I-1$}
        \LineComment{Long-term optimization}
        \If{$\alpha \equiv 0 \pmod{\tau}$}
            \State $(R^i, C^i)_{i = \alpha}^{I-1} \gets$ update forecasts
            \State $(D^i, A^i)_{i = \alpha}^{I-1} \gets \min_{(D^i, A^i)_{i = \alpha}^{I-1}}\sum_{i = \alpha}^{I-1} E^i$  \label{alg:1:lt_optimization}
        \EndIf
        \LineComment{Short-term optimization}
        \State $(D^i, A^i)_{i = \alpha}^{\alpha + \gamma} \gets \min_{(D^i, A^i)_{i = \alpha}^{\alpha + \gamma}} \sum_{i = \alpha}^{I-1} E^i$  \label{alg:1:st_optimization}
        \LineComment{Progress in time}
        \State $\text{execute\_interval}(D^\alpha)$
        \State $R^\alpha, C^\alpha, D^\alpha, A^\alpha \gets$ update past with observed reality
    \EndFor
    \end{algorithmic}
\end{algorithm}

We perform the optimization in two distinct steps:
\begin{enumerate}
    \item A \emph{long-term optimization}, ensures global feasibility, i.e., meeting QoR constraints. It is executed every $\tau$ intervals (e.g., every 24 hours) and solves the MILP for the remainder of the year.
        Any past $D^i$ and $A^i$ are fixed, so we only optimize $(D^i, A^i)_{i = \alpha}^{I-1}$, where $\alpha$ is the current interval.
        Solutions may be approximate due to solver time limits.
    \item A \emph{short-term optimization} adapts to recent forecast updates and improves suboptimal long-term decisions.
    It is executed every interval and solves the problem within a fixed horizon\footnote{We found that the validity period length $\gamma$ is a good indicator for horizon length, although it can be shorter or longer (as future values of $A^i$ are fixed by the long-term optimization). In Line~\ref{alg:1:st_optimization} we assume $\gamma$ as the horizon.}.
        As we only optimize over $(D^i, A^i)_{i = \alpha}^{\alpha + \gamma}$, this optimization has a significantly smaller search space and is more likely to find an optimal solution quickly.
        If no solution is found, we select an $A^i$ such that $\text{QoR}(\alpha,\omega) = 1$, and determine the minimal $D^i$ that satisfies Equation~(\ref{eq:sufficient_resources}).
\end{enumerate}

After optimization, the determined deployment $D^\alpha$ is provisioned for the current interval.
As short-term load and carbon intensity predictions are usually very precise, and the provisioning of new nodes can take significant time (Microsoft reports 6-8 minutes for creating a new LLM inference instance~\cite{stojkovic2024dynamollm}) we assume that there are usually no rapid auto-scaling decisions within an interval.
After the interval has passed, we update the actually observed $D^\alpha$ and $A^\alpha$, alongside the observed number of requests and carbon intensity.

\subsection{Automatic QoR Adaptation under an Annual Budget}
\label{sec:automatic_qor_budget}

The optimization problem in Section~\ref{sec:online} minimizes emissions under a given $\QoRtarget$.
However, especially for best-effort users, an alternative objective is to provide the \emph{highest possible} service quality while staying within an annual carbon budget $B$.
This setting is relevant when operators (or tenants) face internal sustainability targets, reporting requirements, or carbon pricing that effectively turns emissions into a constrained resource.

We model this by making $\QoRtarget$ a decision variable and optimizing
\begin{align}
    \max \quad
    \label{eq:objective_budget} & \QoRtarget \\
    \text{s.t.}\quad
    & (\ref{eq:all_requests_allocated}) - (\ref{eq:sufficient_resources}) \nonumber \\
    \label{eq:below_budget} & \sum_{i=\alpha}^\omega E^i \leq B \\
    \label{eq:qor_target_budget} & \QoRtarget \leq \min \left\{ \text{QoR}(i, i + \gamma)\right\}_{i = \alpha}^{\omega}.
\end{align}

In practice, when re-optimizing at time $\alpha$, a part of the budget has already been spent.
Let $E^0,\dots,E^{\alpha-1}$ be the (observed) emissions of past intervals.
Then the remaining budget is
\begin{align}
    B_\text{rem}(\alpha) = B - \sum_{i=0}^{\alpha-1} E^i,
\end{align}
and the long-term optimization step is solved over the remaining horizon with $\sum_{i=\alpha}^{\omega} E^i \le B_\text{rem}(\alpha)$ while keeping all past decisions fixed.
In the multi-horizon optimization (Algorithm~\ref{alg:1}), we use \eqref{eq:objective_budget} as the objective for the long-term optimization step.
This enables periodic updates of $\QoRtarget$ based on forecast updates and already-incurred emissions, while the short-term step continues to refine per-interval decisions within a smaller horizon.

\section{Experiments}

We evaluate our approach by simulating an LLM inference service over the year 2023 using 8 real and synthetic request traces in 10 different regions.
All experiments were provided 16 cores of an HPC cluster, using Gurobi~\cite{gurobi} for optimization.

\parag{Scenario}
We investigate an LLM inference scenario, where a single user group is served by a LLaMA 3.1~\cite{meta2024llama3} model through the inference and serving engine vLLM~\cite{kwon2023vLLM}.
Users can be fully served by LLaMA 3.1 8B ($\text{QoR}=0$), LLaMA 3.1 70B ($\text{QoR}=1$), or a fraction of both models ($0 < \text{QoR} < 1$).
We consider $\mathcal{M} = \{\text{EC2 p4d.24xlarge}\}$, currently the only instance type with high-performance GPUs available on AWS, and define $p_m^\text{attr} = 3781.8\,$W and $C_m^\text{emb} = 135.3$\,gCO$_2$ based on~\cite{teads2024aws_footprint_estimator}.
We model the machine performance $K$ using recent benchmarks of vLLM on a EC2 p4d.24xlarge for inference, which state a throughput of 11.57 requests per second for LLaMA 3.1 8B and 5.05 requests per second for LLaMA 3.1 70B~\cite{vLLM_benchmark}.

\parag{Request traces}
We require 4 years of hourly request traces: 3 for fitting forecasting models and 1 for the analysis itself.
As such data are not publicly available for LLM inference services, we evaluate our method across a diverse set of other request traces:

\begin{itemize}
    \item \textbf{Artificial}: A \emph{Static} trace that assumes a constant stream of hourly requests, and a \emph{Random} trace, where we sample hourly requests from a normal distribution.
    \item \textbf{Human-Generated}: Two real-world traces from unrelated domains, namely the English and German Wikipedia Pageview statistics~\cite{wikipeadia_data}, \emph{Wiki~(en)} and \emph{Wiki~(de)}, and the hourly taxi trips in New York City~\cite{tlc_trip_record_data}, called \emph{Taxi}.
    \item \textbf{Synthetic}: Three traces generated from models fitted to Google's Borg cluster cells~\cite{google_cluster_traces}, called \emph{Cell B, D, and F}.
\end{itemize}
All request trace datasets and our forecast methodology are explained in detail in Appendix~\ref{sec:datasets}.

\parag{Carbon intensity traces}
We evaluate the potential for ten globally distributed regions:
\begin{itemize}
    \item In Europe, we consider Germany (DE), Spain (ES), Netherlands (NL), Poland (PL), and Sweden (SE)
    \item In the US, we consider California (CISO), Texas (ERCOT), New York (NYISO), and the Pennsylvania-Jersey-Maryland Interconnection (PJM)
    \item In Australia, we consider Queensland (AU-QLD)
\end{itemize}
All carbon intensity data was provided by Electricity\-Maps~\cite{electricityMaps}.
We normalized all time zones to align the time of day with the request traces.
Our forecast methodology is explained in Appendix~\ref{sec:appendix_carbon_intensity}.

\subsection{Absolute Emissions}
\label{sec:eval_qor_energy_efficiency}

\begin{figure}[h]
\centering
  \includegraphics[width=.9\columnwidth]{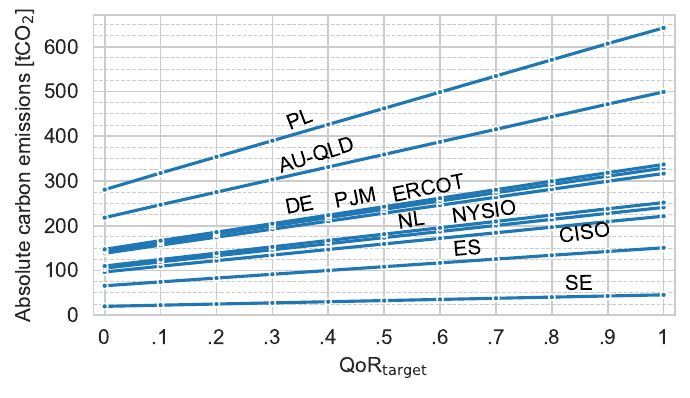}
  \vspace{-5mm}
  \caption{Annual operational carbon emissions for \emph{Wiki~(de)} at different $\QoRtarget$ \emph{without} carbon-aware QoR adaption.}
  \label{fig:absolute_emissions_wiki_en}
\end{figure}

We first assess the absolute annual emissions \emph{without} carbon-aware QoR adaption.
Figure~\ref{fig:absolute_emissions_wiki_en} shows the annual emissions for \emph{Wiki~(de)}.
As expected, emissions scale linearly with $\QoRtarget$, but we observe large differences across regions.
For example, operating the service in Sweden emits almost 27 times less carbon than in Poland.

For reference, ChatGPT, one of the biggest LLM services today, reached over 200 million active users in 2024~\cite{openai2024chatgptusers}.
Assuming just 5 requests per user per day, this amounts to nearly 42 million requests per hour---two orders of magnitude more than our \emph{Wiki~(en)} trace.
Assuming similar performance as Llama 3.1 70B, we estimate that \emph{the annual operational carbon emissions of large-scale LLM service providers already exceed multiple 10,000 tCO$_2$}.

\subsection{Upper-Bound Potential}
\label{sec:eval_upperbound}

To assess the potential of carbon-aware QoR adaptation, we assume perfect forecasts, and solve Equation~\eqref{eq:objective_1} for $\gamma = \{$8~hours, 1~day, 1~week, 1~month$\}$.
Shorter periods show limited potential, as ACI is reported hourly and tends to vary only slowly over time.
Note, that long validity periods can lead to reduced service quality over extended periods, see Appendix~\ref{sec:low_qor_periods}.
We optimize the objective until it is within 0.1\,\% of the optimal solution or after 1 hour. %

Table~\ref{tab:overview} presents the potential for \emph{additional} carbon savings for $\QoRtarget = 0.5$ at $\gamma = 1$\,week.
That is, instead of meeting the 50\,\% target hourly, we ensure that 50\,\% of requests are served by LLaMA 3.1 70B over each week.
We observe that this increased flexibility enables emissions savings of around 8\,\% in many regions.
While request trace patterns exhibit little difference in potential, low-volume traces (\emph{Wiki (de)} and \emph{Taxi}) offer greater savings due to the larger impact of discrete scheduling decisions.
For example, the volume of requests in \emph{Wiki (de)} results in 10--25 active machines at a time, while \emph{Wiki (en)} requires 80--200 active machines.
Note, that the potential depends heavily on $\QoRtarget$ as shown in Appendix~\ref{sec:impact_qor_target}.

\begin{table}
\centering
\caption{Relative upper-bound carbon savings potential.} %
\label{tab:overview}
\resizebox{\columnwidth}{!}{
\begin{tabular}{lcccccccccc}
\toprule
\multirow{2}{*}{\textbf{Region}} & \multicolumn{2}{c}{\textbf{Artificial}} & \multicolumn{3}{c}{\textbf{Human-Generated}} & \multicolumn{3}{c}{\textbf{Synthetic}} & \multirow{2}{*}{\textbf{Mean}} \\
\cmidrule(lr){2-3} \cmidrule(lr){4-6} \cmidrule(lr){7-9}
& \textbf{Static} & \textbf{Rnd.} & \textbf{Wiki} & \textbf{Wiki} & \textbf{Taxi} & \textbf{Cell} & \textbf{Cell} & \textbf{Cell} &  \\
& \textbf{} & \textbf{} & \textbf{(en)} & \textbf{(de)} & \textbf{} & \textbf{B} & \textbf{D} & \textbf{F} &  \\
\midrule
NL     & 8.2 & 8.5 & 7.6 & 10.2 & 11.3 & 7.8 & 7.6 & 8.0 & 8.7±1.3 \\
CISO   & 7.0 & 7.5 & 7.5 & 10.0 & 11.3 & 6.6 & 6.3 & 6.8 & 7.9±1.7 \\
ES     & 7.1 & 7.3 & 6.5 & 9.4 & 10.4 & 6.6 & 6.6 & 6.9 & 7.6±1.4 \\
AU-QLD & 7.9 & 8.4 & 7.1 & 6.1 & 8.8 & 7.6 & 6.9 & 8.0 & 7.6±0.8 \\
DE     & 6.3 & 6.7 & 5.7 & 8.7 & 9.7 & 5.9 & 5.8 & 6.1 & 6.9±1.4 \\
PL     & 4.1 & 4.5 & 3.5 & 6.7 & 7.5 & 3.8 & 3.7 & 4.1 & 4.7±1.4 \\
ERCOT  & 4.2 & 4.6 & 3.6 & 6.3 & 7.4 & 3.9 & 3.7 & 4.1 & 4.7±1.3 \\
SE     & 2.7 & 3.0 & 2.0 & 4.9 & 6.1 & 2.3 & 2.2 & 2.6 & 3.2±1.4 \\
NYISO  & 2.3 & 2.7 & 1.7 & 4.6 & 5.6 & 1.9 & 1.8 & 2.2 & 2.8±1.4 \\
PJM    & 1.8 & 2.3 & 1.3 & 4.1 & 5.5 & 1.6 & 1.4 & 1.8 & 2.5±1.4 \\
\bottomrule
\end{tabular}
}
\end{table}

Figure~\ref{fig:online_results} reports the relative and absolute saving potential for \emph{Wiki (de)} at $\QoRtarget = 0.5$ and different validity periods in gray.
Increasing $\gamma$ to 8 hours yields only limited additional savings---less than 3\,\% across all settings.
This is because average carbon intensity shows limited short-term variation, leaving little room for optimization.
However, starting at $\gamma = 24$\ hours, we see a more notable potential of 5-8\,\% in some regions.
Regional characteristics significantly influence the expected gains:
For instance, while California's carbon intensity primarily fluctuates on a daily basis due to its heavy reliance on solar energy, Germany exhibits more complex patterns with daily, weekly, and seasonal variations~\cite{wiesner2021letswaitawhile}.
As a result, extending $\gamma$ beyond 24 hours has minimal effect in California, while in Germany, larger $\gamma$ values lead to considerable improvements.
In contrast, regions like Sweden, New York, and PJM exhibit little temporal variation in carbon intensity and have therefore a low savings potential of below 5\,\%.

\subsection{Potential under Realistic Conditions}
\label{sec:eval_online}

\begin{figure}[t]
\centering
  \vspace{-1mm}
  \includegraphics[width=\columnwidth]{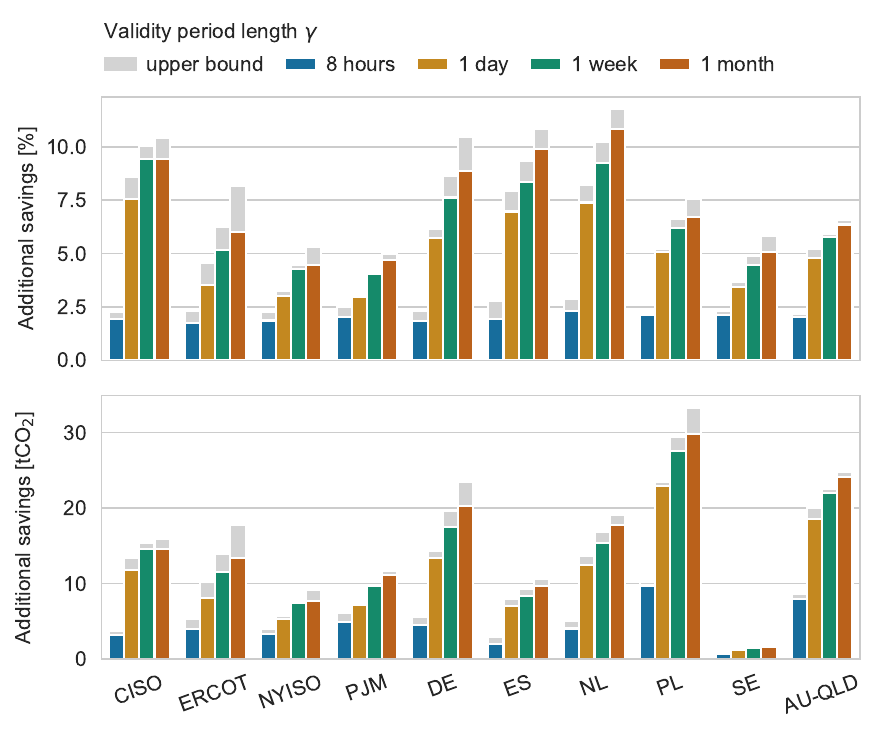}
  \vspace{-7mm}
  \caption{Relative (top) and absolute (bottom) additional carbon savings for different validity periods of our proposed online approach (colored) compared to upper bound potential (gray). Larger validity periods enable higher savings.}
  \label{fig:online_results}
  \vspace{-3mm}
\end{figure}

We evaluate the practicability of carbon-aware QoR adaptation by applying our online approach under real forecasts (see Appendix~\ref{sec:datasets} and \ref{sec:appendix_carbon_intensity}) and approximated MILP solutions:
Long-term optimizations are stopped after 30s; short-term optimizations after 10s.

Figure~\ref{fig:online_results} shows the gap to the upper-bound performance.
Across all experiments, savings under realistic conditions reached $82\pm6$\,\% of the upper-bound potential.
Notably, even in regions with highly unpredictable request patterns (\emph{Random} and \emph{Cell B, D and F}), the online optimization does not show a significant performance drop.

\subsection{Automatic QoR Adaptation under an Annual Budget}
\label{sec:eval_budget}

Lastly, we evaluate the automatic adaptation of $\QoRtarget$ introduced in Section~\ref{sec:automatic_qor_budget} on \emph{Wiki (en)} in three exemplary regions.
For each region, we define an annual budget based on the resulting emissions of the corresponding upper-bound experiment for $\gamma = 24$ hours and $\QoRtarget = 0.5$ (Section~\ref{sec:eval_upperbound}).
Consequently, under perfect forecasts, an optimal solution would maintain $\QoRtarget \approx 0.5$ throughout the year.

\parag{Baselines}
We compare our approach to two greedy baselines that use the same forecasts as our method:
\begin{enumerate}
    \item \emph{Greedy (constant)} divides the remaining annual budget into equal per-interval budgets and greedily selects the highest feasible per-interval quality given the current forecast. Any remaining or overspent budget is evenly distributed across the remaining intervals.
    \item \emph{Greedy (weighted)} follows the same approach but weights the per-interval budgets by forecasted load and carbon intensity. Weightings are updated whenever forecasts are updated.
\end{enumerate}

\begin{figure}[h]
\centering
  \includegraphics[width=\columnwidth]{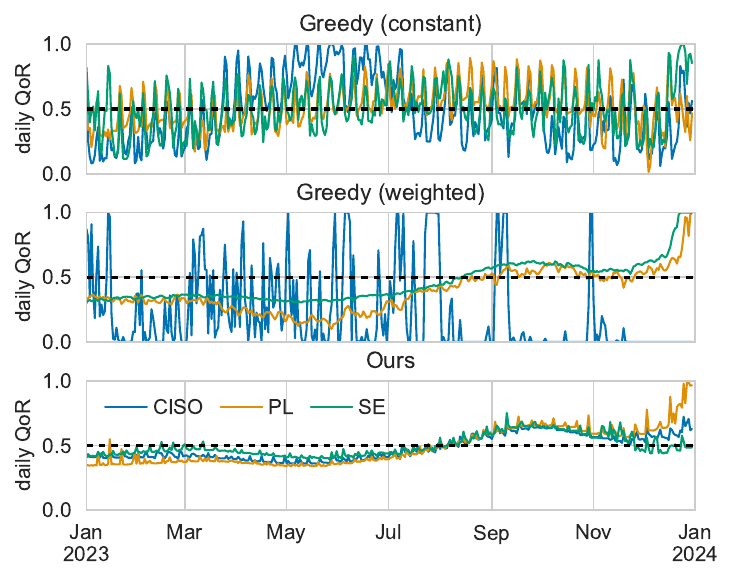}
  \vspace{-7mm}
  \caption{Daily QoR for California (CISO), Texas (ERCOT), and Poland (PL) under an annual carbon budget. Automatic adaptation provides consistent QoR throughout the year.}
  \label{fig:adaptive_1}
  \vspace{-3mm}
\end{figure}

Figure~\ref{fig:adaptive_1} shows the resulting daily QoR under realistic forecasts.
Since \emph{Greedy (constant)} does not account for fluctuations in requests or carbon intensity, the provided QoR varies significantly throughout the year.
\emph{Greedy (weighted)} is more stable but can still drift toward extreme QoR values toward the end of the year as forecast errors accumulate.
In contrast, periodic long-term re-optimizations allow our approach to adjust $\QoRtarget$ over time, yielding more consistent QoR while staying within the annual budget.

\section{Conclusion}

In this paper, we presented a first analysis of carbon-aware QoR adaptation, a strategy to reduce operational carbon emissions of cloud services without relying on geo-distributed load balancing or delay-tolerant workloads.
Our results demonstrate that adjusting service quality in response to carbon intensity can reduce emissions by up to 10\,\% beyond savings related to increased energy efficiency.

To cover a wider range of realistic settings, future work should extend our analysis to scenarios with heterogeneous machines, multiple quality tiers, and diverse user groups. It is also important to examine user-side effects, such as whether low QoR might trigger behavioral responses like repeated requests, potentially offsetting carbon savings.
Additionally, future work should explore the potential of QoR adaptation under sustainability metrics used for short-term demand management, like marginal carbon intensity or excess power during curtailments~\cite{wiesner2025marginal}.
These metrics exhibit higher temporal variability than average carbon intensity, potentially enabling higher savings and the use of shorter validity periods.

\begin{acks}
This research was supported by the German Ministry for Education and Research (BMBF) as BIFOLD---Berlin Institute for the Foundations of Learning and Data (ref. BIFOLD24B), the German Research Foundation (DFG) as FONDA---Foundations of Workflows for Large-Scale Scientific Data Analysis (ref. CRC 1404), and the EU Horizon project EXIGENCE under Grant Agreement No. 101139120.
\end{acks}

\bibliographystyle{ACM-Reference-Format}
\bibliography{bibliography}

\appendix

\section{Nomenclature}
\label{sec:nomenclature}

\begin{table}[h!]
    \centering
    \small
    \caption{Description of symbols. Small letters denote the elements of the corresponding, capitalized matrix.}
    \vspace{-2mm}
    \begin{tabular}{lp{6cm}}
        \toprule
        & \textbf{Input variables} \\
        \midrule
        $\mathcal{U}$ & Set of user groups $u$\\
        $\mathcal{M}$ & Set of machine types $m$\\
        $\mathcal{Q}$ & Tuple of two service quality tiers $q$\\
        $\gamma \in \mathbb{N}$ & Validity period length \\
        $R^i \in \mathbb{R}^{\left|\mathcal{U}\right|}$ & Number of requests $\left(r^i_u\right)$ for user group $u$ during ~$i$. \\
        $C^i \in \mathbb{R}$ & Carbon intensity during $i$. \\
        $P^i \in \mathbb{R}^{\left|\mathcal{Q}\right| \times \left|\mathcal{M}\right|}$ & Power usage $\bigl(p^i_{m,q}\bigl)$ of machine $m$ serving quality $q$ during $i$. \\
        $K \in \mathbb{R}^{\left|\mathcal{Q}\right| \times \left|\mathcal{M}\right|}$ & Requests per interval $\bigl(k_{m,q}\bigl)$ that machine $m$ can serve at quality $q$. \\
        $C_m^\text{emb} \in \mathbb{R}$ & Attributed embodied emissions for running machine $m$ for time $\Delta$. \\
        \midrule
        & \textbf{Decision variables} \\
        \midrule
        $D^i \in \mathbb{N}^{\left|\mathcal{Q}\right| \times \left|\mathcal{M}\right|}$ & Deployment $\bigl(d^i_{m,q}\bigl)$, i.e., the number of machines $m$ running quality $q$ during $i$. \\
         $A^i \in \mathbb{R}^{\left|\mathcal{U}\right| \times \left|\mathcal{Q}\right|}$ & Number of requests $\bigl(a^i_{u,q}\bigl)$ served by quality~$q$ for user group $u$ during $i$. \\
        \bottomrule
    \end{tabular}
    \label{tab:symbols}
\end{table}

\section{Proof of NP-Hardness}
\label{sec:nphard}

We prove that the optimization problem defined in Equations~(\ref{eq:objective_1})--(\ref{eq:min_qos_constr}) is NP-hard via a polynomial-time reduction from the \emph{Bin Packing problem}, which is known to be NP-complete~\cite{garey1979computers}.
Bin Packing is defined as: \emph{Given a set of items with sizes $a_j \in (0,1]$ and an integer $B$, can the items be packed into at most $B$ bins of capacity 1?}

Consider a simplified instance of our problem with a single interval $i$, one user group $u$, one machine type $m$, and one quality tier $q$. Each machine has capacity $k_{m,q}=1$. Requests correspond exactly to Bin Packing items, each with size $a_j$, and the total request volume is $\sum_j a_j$.
In this setting, determining feasibility under Constraints~(\ref{eq:all_requests_allocated}) and~(\ref{eq:sufficient_resources}) is equivalent to solving the Bin Packing problem:
each machine corresponds to a bin, each request to an item, and the capacity constraint ensures that the sum of assigned request sizes per machine does not exceed 1.
Since this special case of our problem is equivalent to Bin Packing, and Bin Packing is NP-complete, it follows that our optimization problem is NP-hard.
\hfill$\square$

\section{Performance Analysis}
\label{sec:performance_analysis}

To illustrate the performance improvements achieved by the proposed multi-horizon optimization, we analyze 240 experiments (all regions, all trace datasets, and $\QoRtarget \in \{0.3, 0.5, 0.7\}$) for the annual-horizon and daily-horizon optimization. These correspond to the long-term and short-term optimization components of our evaluated online approach.

\begin{figure}[h]
\centering
  \includegraphics[width=\columnwidth]{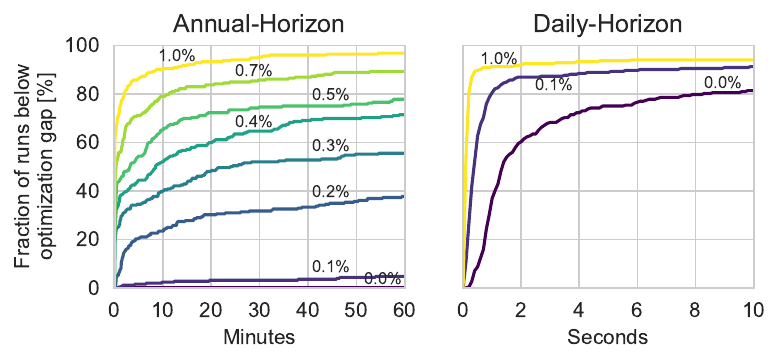}
  \vspace{-8mm}
  \caption{CDF of optimization gap over time.}
  \label{fig:optimization_performance}
\end{figure}

Figure~\ref{fig:optimization_performance} shows a CDF over the fraction of runs below a certain optimization gap over time.
In the annual-hori\-zon optimization, around 10\,\% of all runs have not found a solution within 1\,\% of the optimum within 10 minutes, while no run found an optimal solution within one hour.
In the daily-horizon optimization, the median runtime was only 1.2 seconds and more than 80\,\% of all runs found an optimal solution within 10 seconds.

\section{Request Traces}
\label{sec:datasets}

This section describes all request trace datasets used in the experiments.
Table~\ref{tab:datasets} presents summary statistics.

\begin{table}[h]
\centering
\caption{Request trace and 24-hour forecast statistics.}
\vspace{-3mm}
\label{tab:datasets}
\begin{tabular}{lcccc}
\toprule
& \multicolumn{3}{c}{Dataset statistics ($1 \times 10^6$)} & \multicolumn{1}{c}{Forecast statistics} \\
\cmidrule(r){2-4} \cmidrule(){5-5}
   & {Mean ± Std Dev} & {Min} & {Max} & {MAPE} \\
\midrule
Static     & 1.00 ± 0.00 & 1.00 & 1.00 & 0.0 ± 0.0 \\
Normal     & 1.00 ± 0.34 & 0.00 & 2.36 & 38.6 ± 24.6 \\
Wiki (en)  & 3.38 ± 0.80 & 1.88 & 16.41 & 13.9 ± 8.4 \\
Wiki (de)  & 0.42 ± 0.24 & 0.04 & 1.56 & 32.1 ± 15.3 \\
Taxi       & 0.33 ± 0.14 & 0.04 & 0.71 & 26.5 ± 7.1 \\
Cell B     & 1.94 ± 0.61 & 0.73 & 4.10 & 27.2 ± 13.5 \\
Cell D     & 2.87 ± 0.80 & 1.02 & 7.76 & 22.1 ± 15.8 \\
Cell F     & 1.58 ± 0.41 & 0.87 & 4.32 & 18.2 ± 9.3 \\
\bottomrule
\end{tabular}
\end{table}

\parag{Artificial Datasets}
We artificially create two traces:
\begin{itemize}
    \item \emph{Static} assumes a constant stream of $1 \times 10^6$ hourly requests.
    \item \emph{Random} comprises points randomly sampled from a normal distribution with a mean of $\gamma=1 \times 10^6$ and a standard deviation of $\sigma=0.33 \times 10^6$."
\end{itemize}
During online optimization, we always assume the mean of $1 \times 10^6$ requests per hour as forecasts.

\parag{Real Datasets}
We create three traces based on real data:
\begin{itemize}
    \item \emph{Wiki (en)} is based on the Wikipedia pageview statistics~\cite{wikipeadia_data}, namely all hourly requests to \url{en.wikipedia.org}.
    \item \emph{Wiki (de)} represents all requests to \url{de.wikipedia.org}. The trace differs notably from \emph{Wiki (en)}, as the German Wikipedia is primarily accessed from a single timezone, while the English Wikipedia sees traffic from around the globe.
    \item \emph{Taxi} is based on hourly aggregates of taxi trips in New York City~\cite{tlc_trip_record_data}. We sum over all trip records (Yellow Taxi, Green Taxi, For-Hire Vehicle, High Volume For-Hire Vehicle) and multiply the events by factor 10, to bring them into a similar range with the other datasets.
\end{itemize}

To simulate realistic forecasts, we are using using Prophet~\cite{prophet2017}, a state-of-the-art forecasting model that fits daily, weekly, and annual seasonalities.
In particular, we fit a model every day at midnight using 3 years of historical data to forecast the rest of the year.
The predictions exhibit realistic errors, denoted in Table~\ref{tab:datasets}.

\parag{Synthetic Datasets}
As all real datasets exhibit very periodic patterns, we additionally created three synthetic datasets to demonstrate that carbon-aware QoR adaptation is feasible on highly unpredictable traces.
For this, we aggregated the instance events of the Google cluster traces (version~3)~\cite{google_cluster_traces} individually for the eight available Borg cells.
Data was retrieved via the GoogleSQL query:
\begin{verbatim}
SELECT
  TIMESTAMP_TRUNC(TIMESTAMP_MICROS(time), HOUR) AS hour,
  COUNT(*) AS num_events
FROM `google.com:google-cluster-data.<TABLE>`
-- Exclude invalid values
WHERE time > 0 AND time < 9223372036854775807
GROUP BY hour
ORDER BY hour;
\end{verbatim}
where \verb#<TABLE># is substituted by the name of the cell. For example, for cell A: "\verb#clusterdata\2019\a.instance\events#".

We identified the three traces with the lowest 24-hour autocorrelation coefficients, i.e. the traces that exhibit the least daily seasonality: \emph{Cell B} (0.17), \emph{Cell D} (0.27), and \emph{Cell F} (0.22).
Since the Google cluster dataset only covers one month of data, we fit a DoppelGANger~\cite{lin2019dgan} model for each of the traces, which is a state-of-the-art GAN for timeseries generation, and generate 4 years of synthetic data per cell.
We apply the same forecasting methodology that is used for real datasets.

\section{Carbon Intensity Forecasts}
\label{sec:appendix_carbon_intensity}

\begin{table}[t]
    \centering
    \caption{Carbon intensit forecast MAPE over 96 hours as reported in CarbonCast~\cite{maji2022carboncast}}
    \vspace{-2mm}
    \begin{tabular}{lcccc}
        \toprule
        Region & Day 1 & Day 2 & Day 3 & Day 4 \\
        \midrule
        CISO    & 8.08  & 11.19  & 12.93  & 13.62 \\
        PJM     & 3.69  & 4.93   & 5.87   & 6.67  \\
        ERCOT   & 9.78  & 10.93  & 11.61  & 12.23 \\
        NYISO   & 6.91  & 9.06   & 9.95   & 10.42 \\
        SE      & 4.29  & 5.64   & 6.43   & 6.74  \\
        DE      & 7.81  & 10.69  & 12.80  & 15.55 \\
        PL      & 3.12  & 4.14   & 4.72   & 5.50  \\
        ES      & 10.12 & 16.00  & 19.37  & 21.12 \\
        NL      & 6.06  & 7.87   & 9.08   & 9.99  \\
        AU-QLD  & 3.93  & 3.98   & 4.06   & 5.87  \\
        \bottomrule
    \end{tabular}
    \label{tab:CarbonCast}
\end{table}

For long-term carbon intensity forecasts, we apply the same forecasting methodology that is used for the request trace datasets.
As in \cite{lechowicz2024carbonclipper}, we generate synthetic short-term forecasts of up to 4 days by adding Gaussian noise to the actually observed carbon intensity data to match the MAPEs reported by the state-of-the-art forecasting model CarbonCast~\cite{maji2022carboncast} for each region, see Table~\ref{tab:CarbonCast}.
Short-term forecasts are updated daily at midnight.

\section{Impact of QoR\textsubscript{target}}
\label{sec:impact_qor_target}

\begin{figure}[h]
\centering
  \vspace{-2mm}
  \includegraphics[width=\columnwidth]{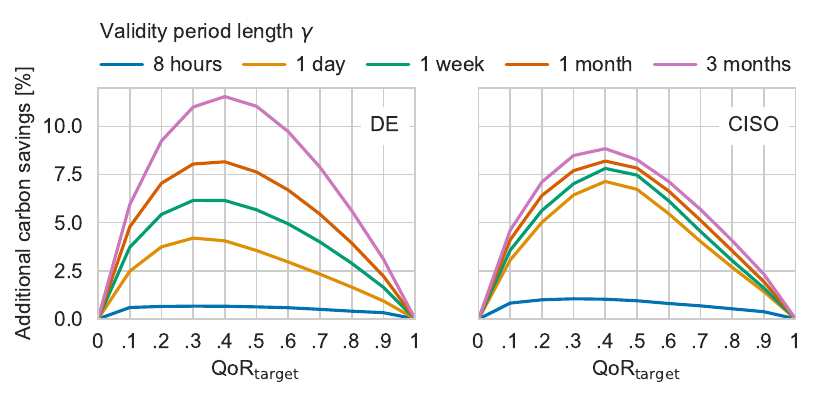}
  \vspace{-9mm}
  \caption{Additional relative savings for different $\QoRtarget$.}
  \label{fig:upper_bound_savings_over_qor}
\end{figure}

Figure~\ref{fig:upper_bound_savings_over_qor} shows the potential for relative savings across different $\QoRtarget$ and validity periods, for \emph{Wiki~(en)} in Germany and California.
At $\QoRtarget = 0$ and $\QoRtarget = 1$, we serve all requests using either LLaMA 3.1 8B or 70B, leaving no flexibility for additional savings.
A $\QoRtarget$ around 0.5 offers the greatest flexibility for carbon-aware QoR adaptation, resulting in the highest potential for carbon savings.

\section{Periods of Low Service Quality}
\label{sec:low_qor_periods}

\begin{figure}[h]
    \centering
    \vspace{-2mm}
    \includegraphics[width=\columnwidth]{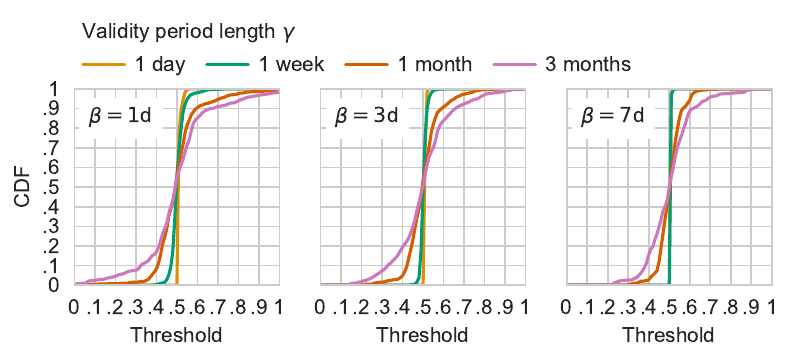}
    \vspace{-7mm}
    \caption{CDF for \emph{Wiki (en)} in Germany at $\QoRtarget = 0.5$: What proportion of time intervals of length~$\beta = \{1, 3, 7\}$ days have a QoR below the threshold?}
    \label{fig:bad_service}
\end{figure}

While increasing $\gamma$ enhances potential for carbon savings, it also increases the probability of prolonged periods with low QoR, illustrated in Figure~\ref{fig:bad_service}.
For example, when optimizing for $\QoRtarget = 0.5$ at $\gamma = 1$ week, no 1-week interval has a QoR below 0.5 (right).
However, 10\,\% of all daily intervals exhibit a $\QoRtarget < 0.48$ (left).
We observe that for $\gamma = 3$ months, users can experience multiple consecutive days of low QoR.
Consequently, large validity periods are not practical in many real-world scenarios.

\section{Equivalence of Provisioning Under Time- vs. Utilization-based Power Attribution}
\label{sec:proof_eq}

We prove that the optimal provisioning decision is the same under (i) utilization-based power attribution (\ref{eq:scope2_powermodel_simple}), where per-machine power depends on resource utilization, and (ii) time-based power attribution, where each active machine is assigned a utilization-independent attributed power draw.
This is important, as utilization-based power attribution introduces a non-linear relationship by multiplying values within $D^{i}$ and $A^i$.
This would turn the MILP into an mixed-integer quadratic program, which is considerably harder to solve.

\begin{assumption}
\label{assumption:1}
$p_{m,q}^\text{max} > p_m^\text{idle} > 0$. \emph{This reflects the basic property that active machines consume more power under load than when idle or turned off.}
\end{assumption}
\begin{assumption}
\label{assumption:2}
$n \leq 1$, \emph{i.e. the power model is concave, which is a common assumption for computing hardware as servers run more efficiently under high load~\cite{xu2017survey_vm_consolidation, gill2018taxonomy_sustainable_computing, barroso2007energy_proportional}.}
\end{assumption}
\begin{assumption}
\label{assumption:3}
$p_{m_1}^\text{idle} < p_{m_2}^\text{idle} \ \Rightarrow \ p_{m_1,q}^\text{max} < p_{m_2,q}^\text{max} \quad \forall m_1, m_2 \in \widehat{\mathcal{M}}$. \emph{If a quality tier can be deployed on different machine types $\widehat{\mathcal{M}}$, we assume a strict ordering of these machines based on their idle and max power usage.}
\end{assumption}

\begin{theorem}
\label{thm:equivalence}
Let Assumptions~\ref{assumption:1},~\ref{assumption:2}, and~\ref{assumption:3} hold. When minimizing \eqref{eq:objective_1} subject to \eqref{eq:all_requests_allocated} -- \eqref{eq:min_qos_constr}, the utilization-based power attribution, 
$$p^i_{m,q} = p_m^\text{idle} + \bigl(p_{m,q}^\text{max} - p_m^\text{idle}\bigl) \cdot \bigl(\text{util}^i_q\bigl)^n,$$ 
can be simplified to the same form as the time-based attribution
$$\tilde{p}^i_{m,q} = p_{m,q}^\text{max},$$
as it always yields an equivalent optimal deployment $D^i$.
\end{theorem}

\begin{lemma}%
\label{lem:no_additional_penalty}
Increasing the utilization of machines does not result in additional penalties, so there is no incentive to deploy more machines of the same type than strictly necessary.
Formally, if Assumptions~\ref{assumption:1} and~\ref{assumption:2} hold, utilization-based attribution $f$ is strictly subadditive
\begin{equation}
    f(x + y) < f(x) + f(y),
\end{equation}
where $x$ and $y$ represent utilization levels $\text{util}^i_q \in (0, 1]$.
\end{lemma}

\begin{proof}[Proof of Lemma~\ref{lem:no_additional_penalty}]
We show that $f$ is strictly subadditive for the two cases $n = 1$ and $n < 1$.

\noindent
\emph{Case 1}: Linear power model ($n = 1$): 
\begin{align}
    f(x + y) &= p_m^\text{idle} + \left(p_{m,q}^\text{max} - p_m^\text{idle}\right) \cdot \left(x + y\right) \\
    f(x + y) &= p_m^\text{idle} + \left(p_{m,q}^\text{max} - p_m^\text{idle}\right) x \notag \\
             &\quad + p_m^\text{idle} + \left(p_{m,q}^\text{max} - p_m^\text{idle}\right) y - p_m^\text{idle}  \\
    f(x + y) &= f(x) + f(y) - p_m^\text{idle}  \\
    f(x + y) &< f(x) + f(y)
\end{align}
\emph{Case 2}: Sub-linear power model ($n < 1$): For a strictly concave function $f$ with $f(0) \geq 0$, $f$ is strictly subadditive on $(0, \infty)$.
\begin{align}
    f(x) &= p_m^\text{idle} + \left(p_{m,q}^\text{max} - p_m^\text{idle}\right) \cdot x^n  \\
    f'(x) &= n \left(p_{m,q}^\text{max} - p_m^\text{idle}\right) \cdot x^{n - 1}  \\
    f''(x) &= \underbrace{n (n-1) \vphantom{\Bigl(}}_{< 1}\ \underbrace{\left(p_{m,q}^\text{max} - p_m^\text{idle}\right)}_{> 1} \cdot \underbrace{x^{n - 2} \vphantom{\Bigl(}}_{> 1} < 0  \\
    f''(x) < 0 &\implies f \text{ is strictly concave}
\end{align}
Additionally, $f(0) = p_{m}^\text{idle} > 0$.
\end{proof}

\begin{proof}[Proof of Theorem~\ref{thm:equivalence}]
For each interval $i$, we define the \emph{minimal deployment} $\check{D}^i$ as any feasible solution that uses the smallest possible number of machines:
\[
  \check{D}^i = \arg\min_{D^i} \Bigl\{
    \sum_{m,q} d^i_{m,q}
    \;\big|\;
    \text{min \eqref{eq:objective_1} s.t. \eqref{eq:all_requests_allocated} -- \eqref{eq:min_qos_constr}}
  \Bigr\}.
\]
We show, that both power attribution models yield $\check{D}^i$:
\begin{itemize}
    \item \emph{Feasibility with fewer machines is impossible:} If a solution $\widehat{D}^i$ has fewer machines than $\check{D}^i$ in at least one dimension, it would violate~\eqref{eq:sufficient_resources}.
    \item \emph{Using extra machines is suboptimal for time-based power attribution:} For $\tilde{p}^i_{m,q}$, any extra machine increases the objective by $p_{m,q}^\text{max}$.
    \item \emph{Using extra machines is suboptimal for utilization-based power attribution:} By Lemma~\ref{lem:no_additional_penalty}, for $p^i_{m,q}$ there is no incentive to deploy extra machines \emph{of the same type}.
    If a quality tier $q$ can be served by multiple machine types $\widehat{\mathcal{M}}$,
    under Assumption~\ref{assumption:3} (a strict ordering of these machines based on their idle and max power usage), if a more efficient machine type $m_1$ is still able to carry extra load, using a less efficient type $m_2$ instead can only increase the objective.  
    Consequently, also across heterogeneous machine types, there is no incentive to add extra machines. 
\end{itemize}

In either case (homogeneous or strictly ordered heterogeneous machines), 
any optimal solution must use exactly the minimal number of machines of each machine type.  
Because both the time-based and the utilization-based models thus select the same 
minimal deployment $\check{D}^i$, they yield equivalent optimal deployments.
\end{proof}

\end{document}